\documentclass[twoside,10pt,openany]{book}
\usepackage{amssymb}
\usepackage{amsmath}

\input{TCILATEX.TEX}
\begin{document}

\author{I.Lubashevsky, V.Gafiychuk}
\title{Mathematical description \\
of heat transfer in living tissue (Part II)}
\maketitle
\tableofcontents

\newpage

\pagestyle{myheadings}

\pagenumbering{arabic}

\part{Appendix}

\markboth{ {\sc \thepart.{ } Appendix}}{}

\chapter{Possible cooperative mechanisms of self-regulation in large natural
hierarchical systems}

\label{ch.App}
\markright
{ {\sc  \thechapter. Possible cooperative mechanisms\ldots}
}

In the previous Chapters we considered in detail the mathematical model for
the vascular network response to variations in the tissue temperature. We
have found that the distribution of the blood temperature over the venous
bed aggregating the information of the cellular tissue state allows the
living tissue to function properly. We think that this property is one of
the general basic characteristics of various natural hierarchical systems.
These systems differ from each other by the specific realization of such a 
\index{synergetic mechanism}synergetic mechanism only. In the present
Chapter generalizing the results obtained above we consider typical examples
of hierarchical systems which can be met in ecology, economy, etc.

\section[What the cooperative self--regulation is from the general point of
view]{What the cooperative self-regulation is \newline
from the general point of view}

\label{sa1}

Large hierarchical systems occurring in nature are characterized by such a
great information the flow that none of their elements can possess whole
information%
\index{information} required of governing the system. Here we want to focus
attention on the fact that in such large hierarchical systems there can be a
cooperative mechanism of regulation which involves an individual response of
each element to the corresponding hierarchical piece of information and
leads to the ideal system functioning due to self-processing%
\index{self-processing} of information.

As a evident example of such systems we may regard living tissue where blood
supplies the cellular tissue with oxygen, nutritious products, etc. and at
the same time withdraws $CO_{2}$, products resulting from living activities
of the cellular tissue. A similar situation takes place in respiratory
systems where oxygen reaches small vessels (capillaries) going through the
hierarchical system of bronchial tubes.

The organization structure of large firms is a clear example of economic
hierarchical systems. Managers of all functions and all levels make up a
management network \cite{app.3}. Roughly speaking, the management network
controls the money flow towards the organization bottom comprising workers
as well as the flow of products in the opposite direction. In performing
technological processes wages paid to workers actually transforms into the
firm products.

The existence of a tremendous amount of goods in market, in contrast to a
relatively small number of raw materials shows that there must be large
hierarchical systems in the market structure. So, it is possible to single
out certain drafts of hierarchical systems inside a given branch of
industry. In this context we note that goods flow on such structures and
after reaching the consumers transform into money flow in the opposite
direction.

Concerning ecological systems we would like to note that they are also
complex in structure, can involve a larger number of ``predator-pray''%
\index{predator-pray levels} levels and are grounded on some basic medium
(for example, plankton) \cite{app.4,ga1,ga5}. Dynamics of ecosystems is
governed by a biomass and energy flow on trophic networks.

In the present Chapter we, first, extend the thermoregulation model stated
above in order to show that the cooperative self--regulation mechanism is
actually based only on the general conservation laws%
\index{conservation law} in systems whose evolution is governed by the
minimum conditions for energy dissipation \cite{ec4,lu0,lu01}. Let us recall
once more the main characteristics that the system under consideration
should possess. Such a system is grounded on a certain distributed basic
medium. This medium is a living continuum which for its activities needs
nutrition or living and a draining system that withdraws products resulting
from its living activities. For this purpose there is a complex transport
network through which transport agent supplies and drains the basic medium.
The architectonics of the transport network should be organized in such a
manner that a flow of a transport agent through the living medium be the
same at each point, all other factors being equal. To meet this condition
the transport network may involve supplying and draining bed of the tree
form. In principle, the two beds can coincide with each other in real space.
Interaction between the transport agent%
\index{transport agent} and the basic medium causes interchange of the
supplied and withdrawn products. The transport agent flow through the basic
medium should keep up the concentrations of these products inside a certain
region called the vital region, which is the aim of regulation.

Due to the transport agent motion being accompanied by energy dissipation it
is necessary that a certain external force to be applied to the system that
affects the overall flow of the transport agent. In living tissue the blood
pressure plays the role of this external force. In order to analyse behavior
of ecological systems it is possible to apply the general principles of
nonequilibrium thermodynamics \cite{app.9}. In economic systems the total
utility function \cite{app.8} seems to play the role of the external force
causing products and a money flow.

In contrast to artificial systems, natural ones are able to adapt to
variations in the environment. Under ordinary conditions the behavior of
natural system as a whole and the individual behavior of its different
elements are likely to follow the strategy of the minimum expenditure rate.
They agree with the minimum entropy production principle%
\index{minimum entropy production principle} stated in nonequilibrium
thermodynamics \cite{app.9}. However, for such complex and nonequilibrium
systems as biological and ecological ones, specification of particular
expressions for the entropy production rate is far from being solved and the
minimum entropy production principle can be applied to these systems at the
phenomenology level only.

It should be noted that, broadly speaking, in systems where functionals
similar to entropy production play the governing role transport agent%
\index{transport agent} flow is actually a free variable. In other words the
transport agent flow distribution over the living system is directly
governed by a certain potential and each element of the system can control
solely its own state rather than transport agent flow through it and other
elements. So, it is likely that in large hierarchical natural systems none
of their elements cannot only possess the total information on the system
but also directly determine transport flow through the network. In these
systems there should be a certain cooperative mechanism of adaptation based
on the individual element control over its own state. The aforesaid allows
us to imagine a living system as one involving two characteristic parts. The
former is living media that consume the products needed for their activity.
The second is a hierarchically organized network flow through which
transport agent delivers these products to each point of the living medium.
The living medium is made up of a large number of elements similar in the
specific function, therefore it may be treated as a continuum. The supplying
network is embedded into this continuum and in its turn involves two parts.
One of them supplies the living medium%
\index{living medium} with the products needed for life activities and the
second withdraws products of life activities from the living medium. Because
the transport agent flow with the life activity products in the draining bed
leaves the living medium and is directed from higher to lower hierarchy
levels it automatically contains all information of the living medium state
distributed over the draining bed. If this information is transmitted to the
supplying bed in such a manner that each its element responds to the
corresponding piece of information then, the supplying bed as a whole will
be able to respond properly to the living medium demand. Feasibility of
synergetic self-regulation in hierarchically living systems based on such
information transmissions and the corresponding response of the supplying
bed is the subject of the present paper.

A real living system should contain a certain element or a subsystem that,
first, causes the cyclic transport agent motion inside the living system,
second, provides the required composition of the products for life activity
in the transport agent flow at the supplying bed entry. The properties and
functioning details of this element or subsystem are beyond the
self-regulation problem. Therefore, in the present model it will be
described as a certain element that joins the stems of the supplying and
draining bed and is characterized by a force causing the transport agent
motion as a whole.

Mathematical model and theory of large hierarchical systems under
consideration and different aspects of the problem of information processing
have been considered in \cite{ec4,lu0,lu01}. It should be noted that the
problem of processing a huge amount of information is also met in
constructing and optimizing large artificial systems similar to transport
networks \cite{app.1,app.2}. Their elements also do not possess the whole
information, however, as a rule, their local interaction with the nearest
neighbors (in space or hierarchy) can directly control the corresponding
mass flow going through them \cite{app.1,app.2}. Elements of the natural
systems either do not possess the whole information or has no capability for
controlling the local flows. They are able to vary their individual
characteristics only, which through the cooperative interaction of all
hierarchy levels leads to the redistribution of a mass flow over the
supplying bed in the proper way.

\section{Generalized mathematical model of living system}

\label{sa2}

Let us consider a system consisting of the distributed living medium $%
\mathcal{M}$ and the transport hierarchical network $\mathcal{N}$ (Fig.~\ref
{Fapp_1}). The transport network involves supplying and draining beds of the
tree form shown in Fig.~\ref{Fapp_1}. by the left and right - hand side
networks and an external element $S$ joining the tree stems. The transport
agent flowing through the former bed supplies the living medium $\mathcal{M}$
with the products needed for life activities. At the same time transport
agent withdraws products of living activities from the living medium through
the draining bed. The medium $\mathcal{M}$ is considered to be a $d$ -
dimensional homogeneous continuum. For any integer number $n$ the medium $%
\mathcal{M}$ can be represented as the union $\mathcal{M}=\cup \mathcal{M}%
_{n}$ of $2^{nd}$ similar disjoint domains $\mathcal{M}_{n}$ whose
characteristic size is $l_{0}2^{-n}$ where $l_{0}$ is the mean size of the
medium $\mathcal{M}$ as a whole. The domains $\{\mathcal{M}_{n}\}$ will be
called fundamental domains of level $n$ and those of the last level $N$ will
be also referred as elementary domains. All the fundamental domains of a
given level form the basic medium $\mathcal{M}$.

\FRAME{ftbpFU}{6.8403cm}{6.7414cm}{0pt}{\Qcb{The stucture of hierarchical
system.}}{\Qlb{Fapp_1}}{Fapp_1}{\special{language "Scientific Word";type
"GRAPHIC";maintain-aspect-ratio TRUE;display "USEDEF";valid_file "F";width
6.8403cm;height 6.7414cm;depth 0pt;original-width 8.4544in;original-height
8.3333in;cropleft "0";croptop "1.0006";cropright "1.0005";cropbottom
"0";filename 'Fapp_1.gif';file-properties "XNPEU";}}

Geometry of both the beds is assumed to be the same, so, we specify it for
the supplying bed only. The stem of this bed splits into $g=2^{d}$ branches
of the first level (Fig.~\ref{Fapp_1}). Each of the first level branches, in
turn, splits into $g$ branches of the second level and so on. The branches
of the last level $(N)$ are directly connected with living medium $\mathcal{M%
}$. The bed is organized in such way that each branch $i$ of a given level $%
n $ supplies a certain domain $\mathcal{M}_{n}^{i}$ as a whole. Each
elementary domain is bound up with the one of last level branches. The last
level number $N$ is assumed to be much larger than unity: $N\gg 1$, and the
length $l_{N}$ may be regarded as an infinitely small spatial scale.\FRAME{%
ftFU}{4.9754cm}{4.7711cm}{0pt}{\Qcb{The fragment of the transport agend flow
network under consideration ($d=2$).}}{\Qlb{Fapp_2}}{Fapp_2}{\special%
{language "Scientific Word";type "GRAPHIC";maintain-aspect-ratio
TRUE;display "USEDEF";valid_file "F";width 4.9754cm;height 4.7711cm;depth
0pt;original-width 8.6853in;original-height 8.3333in;cropleft "0";croptop
"1.0006";cropright "1.0016";cropbottom "0";filename
'Fapp_2.gif';file-properties "XNPEU";}}

For the sake of simplicity we assume that the living medium domain $\mathcal{%
M}$ is a $d$-dimensional cube of edge $l_{0}$ and, corespondent, a
fundamental domain of level $n$ is a cube of edge $l_{0}2^{-n}$. Fragments
of the supplying bed architectonics embedded into the living medium for $d=2$
and $d=3$ is shown in Fig.~\ref{Fapp_2},\ref{Fapp_3}. The transport agent
flow is directed from lower to higher levels of the supplying bed and in the
opposite direction on the draining bed.\FRAME{ftFU}{4.7293cm}{4.6656cm}{0pt}{%
\Qcb{The fragment of the transport agend flow network under consideration ($%
d=3$).}}{\Qlb{Fapp_3}}{Fapp_3}{\special{language "Scientific Word";type
"GRAPHIC";maintain-aspect-ratio TRUE;display "USEDEF";valid_file "F";width
4.7293cm;height 4.6656cm;depth 0pt;original-width 8.4345in;original-height
8.3333in;cropleft "0";croptop "0.9986";cropright "1";cropbottom "0";filename
'Fapp_3.gif';file-properties "XNPEU";}}

In accordances with network architectonics (Fig. ~\ref{Fapp_2},\ref{Fapp_3})
the host transport agent flow goes into and out of the cube $\mathcal{M}_{0}$
through one of its corners. The host transport agent flow of the supplying
bed reaches the cube center $O_{0}$ where it branches out into $2^{d}$ flows
of the first level. Each transport agent flows of the first level reaches a
center $O_{1}$ of one of the $2^{d}$ cubes (called the fundamental domains
of the first level) that compose together the cube $\mathcal{M}_{0}$. At the
centers $\{O_{1}\}$ each of the first level transport agent flows in its
turn branches out into $2^{d}$ second level flows. Then, the flow branching
is continued in a similar way up to level $N\gg 1$.

Transport agent flow going through branches of any level except for the last
level does not interact with the living continuum $\mathcal{M}$. When the
transport agent flow reaches one of the last level branches of the supplying
bed, for example branch $i,$ it uniformly penetrating through the elementary
domain $\mathcal{M}_{N}^{i}$ containing the branch $i$ delivers the products
needed for life activities at each point of the domain $\mathcal{M}_{N}^{i}$%
. At the same time transport agent going through the domain $\mathcal{M}%
_{N}^{i}$ is saturated with the life activity products, thereby it withdraws
the life activity products from the domain $\mathcal{M}_{N}^{i}$ and
transports them through the draining bed out of the living system.

The state of the medium $\mathcal{M}$ will be described by a certain field $%
\theta $ taken to be a measure of the concentration of the life activity
products, which in turn, characterizes the life activity intensity. The
dynamics of the field $\theta $ in the living medium $\mathcal{M}$ is
controlled by volumetric generation due to living activities, disappearance
caused by draining the life medium, and diffusion of life activity products
between elementary domains. Therefore, the field $\theta $ is considered to
evolve according to the equation 
\begin{equation}
\frac{\partial \theta }{\partial t}=D\mathbf{\nabla }^{d}\theta +q-\theta
\eta ,  \label{*2.1}
\end{equation}
where $D$ is the diffusivity, $\mathbf{\nabla }^{d}$ is the $d$-dimensional
Laplace operator, $q$ is the generation rate and $\eta $ is the volumetric
rate of the transport agent flow. Since, branch $i$ of the last level
supplies the elementary domain $\mathcal{M}_{N}^{i}$ of the living medium as
a whole; the current $J_{i}$ of the transport agent in branch $i$ and the
transport agent flow rate $\eta $ in the corresponding domain are related by
the expression 
\begin{equation}
J_{i}=\int\limits_{\mathcal{M}_{N}^{i}}d\mathbf{r}\eta .  \label{*2.2}
\end{equation}

The last term in equation (\ref{*2.1}) implies that due to interaction
between the living medium and transport agent in the last level branches the
concentrations of life activity products $\mathcal{M}$ inside a living
medium elementary domain $\mathcal{M}_{N}^{i}$ and in the transport agent
going through the corresponding last level branch $i$ of the draining bed
became equal to each other. Therefore, first, transport agent going through
the draining bed can be also characterized by variable $\theta $ and the
value $\theta _{i}$ of the variable $\theta $ corresponding to branch $i$
and can be different for various branches due to possible nonuniformities in
life activity of a living medium. Second, the value $\theta $ corresponding
to branch $i$ of the last level and distribution of the field $\theta $ over
the elementary domain $\mathcal{M}_{N}^{i}$ must be related as 
\begin{equation}
J_{i}\theta _{i}=\int\limits_{\mathcal{M}_{N}^{i}}d\mathbf{r}\theta \eta .
\label{*2.3}
\end{equation}

The pattern of the transport agent current flow $\{J_{i}\}$ on the network $%
\mathcal{N}$ obeys the conservation law at branching points. So, for a given
branching point $B$ we can write 
\begin{equation}
\sum\limits_{B}J_{\text{out}}=J_{in}\qquad \text{or \ \ \ \ \ }%
\sum\limits_{B}J_{in}=J_{\text{out}},  \label{*2.4}
\end{equation}
where $J_{in}$ and $J_{\text{out}}$ are the transport agent currents on
branches going in and out of the point $B$ and the sum runs over all the
branches leading to or out this point. The former and latter expressions are
corresponding to the supplying and draining bed, respectively. The life
activity products moves with transport agent flow through the draining bed,
thus the value $J_{i}\theta _{i}$ can be regarded as the current of the
variable $\theta _{i}$ on branch $i$. As the transport agent going through
the draining bed towards its stem does not interact with the living medium,
the pattern $\{J_{i}\theta _{i}\}$ should also obey the conservation law at
the branching points of the draining bed. The latter allows to write for a
given branching point $B$ of the draining bed 
\begin{equation}
\sum\limits_{B}(J_{i}\theta _{i})_{in}=(J_{i}\theta _{i})_{\text{out}}.
\label{*2.5}
\end{equation}
Equation (\ref{*2.1}), among with expressions (\ref{*2.2}) and (\ref{*2.3})
describe distribution of the field $\theta $ over the living medium and its
relation with the transport agent flow through the transport network.
Equation (\ref{*2.4}) and (\ref{*2.5}) reflect the general laws of transport
phenomena in the supplying and draining beds. Now we specify the way how the
system controls the transport agent flow through the network $\mathcal{N}$.
First, it certain effort is needed for the living system to move transport
agent through the supplying and draining beds. Therefore, it would appear
reasonable that the living system tries to minimize the total energy
dissipation%
\index{energy dissipation} due to transport agent motion. In mathematical
terms the total energy dissipation caused by transport agent motion through
the supplying and draining beds as well as through the element $S$ is
represented in the form 
\begin{equation}
\mathcal{D}\{J_{i}\}=%
\frac{1}{2}\sum\limits_{i}R_{i}J_{i}^{2}-J_{0}E_{ext}.  \label{*2.6}
\end{equation}
Here $R_{i}$ is a kinetic coefficient of branch $i$, $E_{ext}$ is the force
that gives rise to transport agent motion through the network $\mathcal{N}$
as a whole, $J_{0}$ is the transport agent flow through the tree stems and
the sum runs over all the branches of the supplying and draining beds.
Therefore the flow pattern on the supplying and draining beds as well as the
coefficient collection $\{R_{i}\}$ must be mirror images of each other
within the reverse flow direction.

Keeping in mind that the system tries to minimize the energy dissipation due
to transport agent motion we postulate that the distribution $\{J_{i}\}$ of
transport agent flow over the network $\mathcal{N}$ corresponds to the
minimum of the functional%
\index{functional} $\mathcal{D}\{J_{i}\}$ subject to conditions (\ref{*2.4}%
). As has been mentioned in Section~\ref{sa1} the distribution $\{\theta
_{i}\}$ can be treated as the aggregated information on the living medium
state. In particular, the variable $\theta _{i}$ corresponding to branch $i$
describes the state of the corresponding fundamental domain $\mathcal{M}%
_{n}^{i}$ as a whole. Therefore, in order to complete the given model we
should specify the response of the transport network $\mathcal{N}$ to
variations in the variables $\{\theta _{i}\}$. Due to supplying and draining
beds being mirror images of each other we may describe this response for the
draining bed only. We assume that for each branch, for example the branch $i$%
, time variations in the transport coefficient $R_{i}$ are directly
controlled by the variable $\theta _{i}$ assigned to this branch.

When time variations of the distribution $\{\theta _{i}\}$ are
quasistationary the coefficient $R_{i}$ corresponding to branch $i$ of level
$n$ is the explicit function of the variable $\theta _{i}$, i.e. $%
R_{i}=R_{n}(\theta _{i})$. The general properties of the $R_{n}(\theta _{i})$
dependence are actually determined by the Fact that the transport agent flow
through the network $\mathcal{N}$ should grow with increase in the life
activity intensity. Indeed, let during the living system adapting to changes
in the environment life activities of the living medium grow in intensity.
The latter leads immediately to increase in the concentration of the life
activity products inside the living medium, and, thus, to increase in the
field $\theta $, causing increase in variables $\theta _{i}$. The higher
life activity intensity, the greater amount of the products needed for life
activities. Therefore, under such conditions transport agent flow through
the network $\mathcal{N}$ must increase too. In other words, increase in the
variable $\theta _{i}$ should give rise to increase of transport agent flow.
As it will be shown below the coefficient $R_{i}$ can be treated as a
resistance of branch $i$ to the transport agent flow through it. Thus, the
function $R_{n}(\theta _{i})$ must be decreasing with respect to the
variable $\theta _{i}$. Taking the aforesaid into account we represent the $%
R_{n}(\theta _{i})$ dependence in terms of 
\begin{equation}
R_{n}(\theta _{i})=R_{n}^{0}\phi (\theta _{i}),  \label{*2.7}
\end{equation}
where $R_{n}^{0}$ is a constant equal to $R_{n}(\theta _{i})$ for $\theta
_{i}=0$ whose value is assumed to be the same for all the branches of one
level and $\phi (\theta _{i})$ is a certain universal function. The
characteristic form of the $\phi (\theta _{i})$ dependence is shown in Fig.~%
\ref{Fapp_4}a., where value $\theta _{c}$ matches the maximum allowable
concentration of the life activity products in the living medium. When the
concentration of life activity products exceeds the allowable value, that is 
$\theta >\theta _{c}$ the living medium goes beyond the vital interval and
loses the capability for adapting. So, for $\theta _{i}>\theta _{c}$ the
function $\phi (\theta _{i})$ has to be practically constant.%
\FRAME{ftbpFU}{7.3279cm}{3.8067cm}{0pt}{\Qcb{Behaviour of the function $%
\protect\phi (x)$ - a and the function $f(x)$ - b.}}{\Qlb{Fapp_4}}{Fapp_4}{%
\special{language "Scientific Word";type "GRAPHIC";maintain-aspect-ratio
TRUE;display "USEDEF";valid_file "F";width 7.3279cm;height 3.8067cm;depth
0pt;original-width 16.121in;original-height 8.3333in;cropleft "0";croptop
"0.9993";cropright "0.9994";cropbottom "0";filename
'Fapp_4.gif';file-properties "XNPEU";}}

For the transport network $\mathcal{N}$ to be able to respond properly to
local variations in life activities of the living medium distribution of the
transport agent flow over the network $\mathcal{N}$ it should be controlled
by branches of all the levels. This condition, as it will be seen below,
allows us to represent the dependence of the value $R_{n}^{0}$ on the level
number $n$ in the form
\begin{equation}
R_{n}^{0}=R_{0}2^{dn}\rho (n),  \label{*2.8}
\end{equation}
where $R_{0}$ is a constant equal to $R_{n}^{0}|_{n=0}$ and $\rho (n)$ is a
smooth function of $n$ such that $\rho (0)=1$ and formally $\rho
(n)\rightarrow 0$ as $n\rightarrow \infty $.

Concluding description of the system response we also take into account the
possible time delay of the branch response to variations in the variables $%
\{\theta _{i}\}$ and represent the evolution equation for the kinetic
coefficient $R_{i}$ of branch $i$ belonging to level $n$ as 
\begin{equation}
\tau _{n}\frac{dR_{i}}{dt}+(R_{i}-R_{n}^{0})f\left( \frac{R_{i}}{R_{n}^{0}}%
\right) =-\frac{\theta }{\theta _{c}}R_{n}^{0},  \label{*2.9}
\end{equation}
where $\tau _{n}$ is the time delay of branch $i$ response. The value $\tau
_{n}$ is assumed to depend on the level number only. We have written the
kinetic coefficient equation in such form that the term $\frac{\theta }{%
\theta _{c}}$ can be regarded a dimensionless signal which is generated by
transport agent receptors of the draining bed and is sent to the
corresponding branches of the supplying bed. As it follows from (\ref{*2.7})
and (\ref{*2.9}) the function $f(x)$and $\phi (x)$ are related by expression 
\begin{equation}
\lbrack 1-\phi (x)]f[\phi (x)]=x.  \label{*3.10}
\end{equation}
The behavior of the function $f(x)$ is displayed in Fig.~\ref{Fapp_4}b. We
introduced the function $f(x)$ related with function $\phi (x)$ only for
convenience of the consideration of the linear model for branch response.

\section{Synergetic self--regulation%
\index{Synergetic self--regulation}}

\label{sa3}

In this Section we analyse the generalized model stated above. In order to
study dynamics of the field $\theta $ in the living medium $\mathcal{M}$ we
should know the relation between the fields $\eta $ and $\theta $. The
latter problem can be solved if we find the pattern $\{J_{i}\}$ of the
transport agent flow on the network $\mathcal{N}$ depending on the two
fields $\theta $ and $\eta $. Indeed, in this case expressions (\ref{*2.2}),
(\ref{*2.3}) specifying the interaction between the living medium $\mathcal{M%
}$ and transport agent flow through the last level branches immediately
leads us to the desired relation $\eta \{\theta \}$.

Let us find the extremal equations%
\index{extremal equations} for functional (\ref{*2.6}). Following the
Lagrangian multiplier%
\index{Lagrangian multiplier} method we reduce the extremum problem for
functional (\ref{*2.6}) subject to conditions (\ref{*2.4}) to finding the
extremal equations of the following functional%
\index{functional} 
\begin{equation}
\mathcal{D}_{L}\{J_{i}\}=%
\frac{1}{2}\sum_{i}R_{i}J_{i}^{2}-E_{ext}J_{0}+\sum\nolimits_{j}^{B}P^{j}%
\left( \sum_{B}(J_{i})_{in}-\sum_{B}(J_{i})_{\text{out}}\right) ,
\label{*3.1}
\end{equation}
where $\{P^{j}\}$ are the Lagrangian multipliers ascribed to each branching
point $B$ of the network $\mathcal{N}$, the sum $\sum_{j}^{B}$ runs over all
the branching points $\{B_{j}\}$, and the symbols $\sum_{B}(J_{i})_{in},%
\sum_{B}(J_{i})_{\text{out}}$ stand for the sum over all the branches going
in or out of a given branching point $B$. (The direction of the motion on
the network $\mathcal{N}$ is chosen to coincide with the transport agent
flow direction.) From the conditions $\partial \mathcal{D}_{L}/\partial
J_{i}=0$ we obtain the desired extremal equations for the branches 
\begin{equation}
J_{i}R_{i}=P_{in}^{i}-P_{\text{out}}  \label{*3.2}
\end{equation}
and 
\begin{equation}
P_{in}-P_{\text{out}}=E_{ext}.  \label{*3.3}
\end{equation}
Here the multipliers $P_{in}^{i}$ and $P_{\text{out}}$ correspond to
branching points that the given branch $i$ goes in and out of, $P_{in}$ and $%
P_{\text{out}}$ are the multipliers ascribed to the input and output of the
transport network. It should be noted that equations (\ref{*3.2}),(\ref{*3.3}%
) together with equation (\ref{*2.4}) are actually made up of the Kirchhoff
equations for the network $\mathcal{N}$ where the variables $\{P^{j}\}$ play
the role of potentials at the branching points $\{B^{j}\}$ causing the
transport agent flow through the corresponding branches.

In this way we have reduced practically one--to--one the analysis of
self-regulation problem to that considered in Chapters 11 and 12 ~\cite{parti}
for living tissue. The latter allows us to make use of the results
obtained there and to go directly to the final results for the generalized
model. For the sake of simplicity we consider only the case when the
supplying and draining beds respond ideally, that is when the function $\phi
(\theta )$ is given by the formula
\begin{equation}
\phi ^{\text{id}}(\theta )=\left\{
\begin{array}{ccc}
1-\theta /\theta _{c} & \quad \text{if}\quad & 0<\theta <\theta _{c} \\
0 & \text{if} & \theta >\theta _{c}
\end{array}
\right. ..  \label{*4.1}
\end{equation}
In this case the rate of the transport agent flow will be actually
determined by formula:
\begin{equation}
\tau \frac{\partial j}{\partial t}+j\left[ 1-\beta _{cc}\frac{T-T_{a}}{%
\Delta }\right] =j_{0}\;.  \label{11.12}
\end{equation}
then, rewriting it we get
\begin{equation}
\tau \dot{\eta}+\eta \left( 1-\frac{\theta }{\theta _{c}}\right) =\eta _{0}.
\label{*4.3.1}
\end{equation}
Formula (\ref{*4.3.1}) is the desired equation governing the ideal response
of the transport network to changes in living medium activities. It is of
the local form, in other words, the transport agent flow rate $\eta $ is
determined by variations in the field $\theta $ at the same point only.
Thus, the cooperative response of all the network branches is so
self-consistent that the transport network delivers products needed for life
only to the living medium point that ``asks'' the network for this. In
advertent changes in delivery of products for life activity does not occur
in this case, i.e. the transport network ``works'' ideally.

The other peculiar property of the ideal transport network is the fact that
the variable $\theta$ cannot go beyond the vital interval $[0, \theta_c]$
for a long time. The latter follows from unbounded increase in the transport
agent flow rate the variable $\eta$ as $\theta$ tends to $\theta_c$.

Now we discuss in detail how the ideal regulation occurs. Let us assume
that, for example, in the domain $Q$ (Fig.~\ref{Fapp_6}) the variable $%
\theta $ exceeds its normal value due to the balance in living activities
being disturbed. In order to smother the increase in the variable $\theta $
the system should increase the transport agent flow rate $\eta $ in the
domain $Q$. The system responds by decreasing the kinetic coefficients $%
\{R_{i}\}_{\mathcal{P}}$ along the whole path $\mathcal{P}$ on the network $%
\mathcal{N}$ that leads from the stem of the supplying bed to the domain $%
\mathcal{M}$ and, then, from this domain to the draining bed stem (Fig.~\ref
{Fapp_6}). Information required of this system behavior is delivered by the
distribution of the variables $\{\theta _{i}\}_{\mathcal{P}}$ over the path $%
\mathcal{P}$ on the draining bed. In fact, increase of the variable $\theta $
in the domain $Q$ must give rise to the corresponding the increase of all
the variables $\{\theta _{i}\}_{\mathcal{P}}$. The increase in $\theta _{i}$
leads to decrease in the coefficient $R_{i}$. This relation between the
field $\theta $ and kinetic coefficients $\{R_{i}\}$ of the network $%
\mathcal{N}$ is the essence of the self-regulation process in the active
hierarchical system under consideration. \FRAME{ftbpFU}{2.994in}{1.471in}{0pt%
}{\Qcb{Schematic representation of the cooperative mechanism of ideal
self-regulation.}}{\Qlb{Fapp_6}}{Fapp_6}{\special{language "Scientific
Word";type "GRAPHIC";maintain-aspect-ratio TRUE;display "USEDEF";valid_file
"F";width 2.994in;height 1.471in;depth 0pt;original-width
11.8151in;original-height 5.7778in;cropleft "0";croptop "1";cropright
"1";cropbottom "0";filename 'Fapp_6.gif';file-properties "XNPEU";}}%
Variations in the field $\theta $ located in the domain $Q$ can, in
principle, cause an alteration of the transport agent flow rate at the
exterior points due to the flow redistribution over the network $\mathcal{N}$%
. However, in contrast to the points of the domain $Q$, at the exterior
points, for example, at points of the domain $Q^{\prime }$ (Fig.~\ref{Fapp_6}%
), the decrease of different kinetic coefficients belonging to the
collection $\{\Lambda _{i}\}_{\mathcal{P}}$ gives rise to variations in the
field $\eta $ different in sign. Fig.~\ref{Fapp_6} schematically shows the
sign of this effect for different branches of the path $\mathcal{P}$. In the
given model due to the specific forms of function (\ref{*4.3.1}) and the
right-hand side of Eq.(\ref{*4.3.1}) the net effect is reduced to zero. For
other forms of these functions the second term in Eq.(\ref{*4.3.1}) will be
of nonlocal form, i.e. time variations in the field $\eta $ at a given point
will be also determined by values of the variable $\theta $ at other points.

The way in which we got the local equation of the self--regulation%
\index{self--regulation} could give the impression that for ideal
self--regulation to be the case the transport network must be of a regular
geometry with respect to branching and embedding into the space. So let us
now show that the ideality of self--regulation holds for transport network
of the arbitrary architectonics, at least, when the transient processes can
be ignored. By way of example, we consider the transport network $\mathcal{N}
$ shown in Fig.~\ref{Fapp40}.
\FRAME{ftbpFU}{4.6788cm}{6.1571cm}{0pt}{\Qcb{Living medium with irregular
network.}}{\Qlb{Fapp40}}{Fapp40}{\special{language "Scientific Word";type
"GRAPHIC";maintain-aspect-ratio TRUE;display "USEDEF";valid_file "F";width
4.6788cm;height 6.1571cm;depth 0pt;original-width 6.3053in;original-height
8.3333in;cropleft "0";croptop "1";cropright "1.0014";cropbottom "0";filename
'Fapp40.gif';file-properties "XNPEU";}}Let at the initial time the
distribution of the fields $\tilde{\theta}(\mathbf{r})$ and $\tilde{\eta}(%
\mathbf{r})$ over the medium $\mathcal{M}$ be predetermined. On the network $%
\mathcal{N}$ the patterns $\{\tilde{\theta}_{i}\}$, $\{\tilde{J}_{i}\}$
correspond to these fields. In this case by virtue of expression~(\ref{*3.2}%
) we can ascribe the potentials $\{\tilde{P}^{i}\}$ to the branching points
of the network $\mathcal{N}$. Due to the supplying and draining beds being
symmetrical all the terminal points of the last level branches are
characterized by the same potential $P^{m}$. First, we will show that the
transport network responds to disturbances in the living medium activities
in such a manner that only the transport agent flows $\{J_{i}\}$ vary
whereas the potentials $\{\tilde{P}^{i}\}$ are constants. For this purpose
let us consider any one branching point of the last level, e.g. the
branching point \emph{B }(Fig.~\ref{Fapp40}). Suppose that the state of the
living medium $\mathcal{M}$ has changed. Assuming the potentials $\{\tilde{P}%
^{i}\}$ to be constant and taking into account~(\ref{*2.4}), (\ref{*2.5}),
and (\ref{*3.2}) we write for all the branches of this node 
\begin{equation}
J_{0}=\sum_{i=1}^{3}J_{i},  \label{app100}
\end{equation}
\begin{equation}
J_{0}\theta _{0}=\sum_{i=1}^{3}J_{i}\theta _{i},  \label{app101}
\end{equation}
\begin{equation}
J_{0}-J_{0}\frac{\theta _{0}}{\theta _{c}}=\frac{\tilde{P}^{B}-\tilde{P}%
^{B^{\prime }}}{R_{0}},  \label{app102}
\end{equation}
and for $i=1,2,3$%
\begin{equation}
J_{i}-J_{i}\frac{\theta _{i}}{\theta _{c}}=\frac{\tilde{P}^{m}-\tilde{P}^{B}%
}{R_{i}}.  \label{app103}
\end{equation}
Here $\{J_{i}\}$ and $\{\theta _{i}\}$ are the new values of the agent flows
and the variable $\theta $ in the branches $i=1,2,3$ and the branch $0$
going into and out of the point \emph{B}. The values $\theta _{1},\theta
_{2},\theta _{3}$ are directly determined by the mean values of the field $%
\theta $ in the domains $\mathcal{M}_{1},\mathcal{M}_{2},\mathcal{M}_{3}$ so
the values $\theta _{1},\theta _{2},\theta _{3}$ should be regarded as given
constants in finding the pattern $\{J_{i}\}$.

For this problem the set of independent variables involves five values $%
J_{1},J_{2},J_{3,}$ $J_{0,}$and $\theta _{0}$, whereas, the number of
equations is equal to six. However, these equations are linearly dependent.
Indeed, summing equations~(\ref{app103}) and taking into account (\ref
{app100}), (\ref{app101}) we obtain 
\begin{equation}
J_{0}-J_{0}\frac{\theta _{0}}{\theta _{c}}=\sum_{i=1}^{3}\frac{\tilde{P}^{m}-%
\tilde{P}^{B}}{R_{i}}.  \label{app104}
\end{equation}
Equations~(\ref{app102}), (\ref{app104}) coincide with each other if 
\begin{equation}
\frac{\tilde{P}^{B}-\tilde{P}^{B^{\prime }}}{R_{0}}=\sum_{i=1}^{3}\frac{%
\tilde{P}^{m}-\tilde{P}^{B}}{R_{i}}.  \label{app105}
\end{equation}
The latter equality is fulfilled because the values $\{\tilde{P}^{i}\}$ have
been established by the initial distribution of transport agent flows $\{%
\tilde{J}_{i}\}$. Therefore the system of equations~(\ref{app100})--(\ref
{app103}) admits a solution.

For the branching point \emph{B}$^{\prime }$\emph{\ }the branch $0$ plays
the same role as branches $i=1,2,3$ for the point \emph{B}, in particular,
the value $\theta _{0}$ should be treated as a predetermined parameter.
Replicating these speculations practically one--to--one with respect to the
point \emph{B}$^{\prime }$ and then, going towards the stem we show that the
total system of equations governing the transport agent flow on the network 
\emph{N} as a whole admits the solution with $\{P^{i}\}=\{\tilde{P}^{i}\}$.
Whence, it follows that variations in the transport agent flow $J_{i}$
through a branch $i$ of the last level is determined by the value of $\theta
_{i}$ only which is equal to the mean value of the field $\theta $ over the
corresponding domain $\mathcal{M}_{i}$. The latter is responsible for the
local relationship between the fields $\theta $ and $\eta $.

The existence of the local relation between the fields $\theta $ and $\eta $
in such a simple form is a surprise because this relationship does not
contain directly information on the complex geometry of the hierarchical
network involving a tremendous number of elements. This fact is actually the
main result of the present Section.

Concerning real systems in nature the obtained results enable one to study
their evolution and behavior from two different standpoints individually.
One problem is the analysis of architectonics of real natural systems and
the product transport flow as well as information flow in them. Another is
connected with the individual properties and behavior of system elements.
From the standpoint of a living system the adaptation transport network and
its elements differ in purpose. The transport network forms required flow of
products and information that is not directly controlled by system elements.
The system elements responding to a local information adapt individually.

The existence of a large hierarchical systems in nature and their capability
for adapting to changes in the environment points the fact that they are
organized and function approximately ideally. Therefore, the general
principles of hierarchical system functioning found in the present paper are
likely to be useful in the analysis of their function disturbance and
adaptation.

In nature, for example, in living tissue in order to reduce this effect of
the nonideality arterial and venous beds contain a system of anastomoses,
i.e. vessels joining arteries or veins of the same level. So, architectonics
of large natural systems is organized in such a manner that their
functioning has to be ideal as perfect as possible.

For different natural systems the variables $\theta $ and $\eta $ are
distinctive in meaning. For example, for living tissue the variable $\theta $
can be treated as the concentration of carbon dioxide, or the tissue
temperature, the variable $\eta $ is the blood flow rate, and the blood
pressure plays the role of the external potential $E_{ext}$ and $d=3$. A
similar situation takes place in respiratory systems, where the variable $%
\theta $ is the oxygen concentration. In economic systems the quantity $\eta 
$ is the flow rate of goods and the variable $\theta $ is the price. In this
case the external potential $E_{ext}$ is likely to be treated as the total
utility function of the production process in a certain industry. In models
for organization and functioning of firms the basic medium is the firm
bottom comprising workers, the variables $\eta $ and $\theta $ are
quantities proportional to the wages of workers and amount of products,
respectively. Concerning ecological systems the variables $\eta $ and $%
\theta $ seem may be regarded as the rate of a biomass flow and the energy
stored up in a biomass unit.

\section{On hierarchical structures arising spontaneously in markets with a
perfect competition}

\label{sa4}

In real economic systems%
\index{economic systems} there must be a certain mechanism that informs
people what they should produce and in what amount, what work should be
performed for this production etc. Broadly speaking, there are two
essentially different mechanisms governing the economic life. One of them is
based on the state hierarchical system, where the behavior of each economic
agent is directly controlled by one of the higher rank \cite{ec1}. According
to another mechanism the correlation in people behavior is grounded on the
spontaneous order which arises through information obtained by individuals
in the interaction with their local economic circumstances. The latter
mechanism is actually the core of real markets%
\index{market} \cite{ec1,ec2,ec3}.

In the present Section we pay attention to the fact that the market can also
contain hierarchical structures arising spontaneously where each of their
elements responds to the corresponding piece of information, solves its
individual problem, e.g. ``maximize'' its own profit%
\index{profit}, which, nevertheless, leads to the perfect functioning of the
system as a whole. The ideality in the behavior of these systems is caused
by self-processing of information at each hierarchy level, i.e. by a
synergetic mechanism of self-regulation \cite{ec4}.

Let us, first, discuss the reason and the place where such hierarchical
structures come into existence.

The existence of a tremendous amount of goods%
\index{good} in market, in contrast to a relatively small number of raw
materials as well as to the producer specialization shows that there must be
a highly complex network which links different producers with one another,
transforms raw materials and, finally, supplies consumers with the goods
required. The structure of such a network is schematically illustrated in
Fig.~\ref{Fapp_1m}. A branch of the given network (e.g., branch $i$) joining
the nearest nodes represents a collection of producing firms that can be
regarded as identical from the standpoint of their input and output. The
firms are linked with one another by sell-buying processes, the output of
firms belonging to higher ranks being input for firms of lower ranks. The
nodes specify these sell-buying interactions. Firms at the last level sell
directly to ultimate consumers, supplying the latter with different types of
goods. In other words, the input and output of different firms form in the
given network a material flow $\{X_{i}\}$ going from raw material stems to
the consumer medium.%
\FRAME{ftbpFU}{2.2191in}{2.9378in}{0pt}{\Qcb{Schematic representation of the
market network.}}{\Qlb{Fapp_1m}}{Fapp_1m}{\special{language "Scientific
Word";type "GRAPHIC";maintain-aspect-ratio TRUE;display "USEDEF";valid_file
"F";width 2.2191in;height 2.9378in;depth 0pt;original-width
6.2777in;original-height 8.3333in;cropleft "0";croptop "1";cropright
"1";cropbottom "0";filename 'Fapp_1m.gif';file-properties "XNPEU";}}It
should be noted that this sell-buying interaction singles out a certain
economic system under consideration that, on one hand, involves a great
number of participants and exhibits a general property typical for all
markets, and, on the other hand, is a small part of the whole market society
producing goods of a certain type. For example, steel, food, clothing
industries may be regarded as such microeconomic markets%
\index{microeconomic market}. In living organisms, regional vascular network
of different organs play the same role as microeconomic markets do in the
human society. In these terms the material flow corresponds to the blood
flow in vessels and the consumer medium is related to the cellular tissue.

Particular interconnections between different firms can occur and disappear
during formation and evolution of the market under consideration. These
interactions are governed by trade with each other. The latter process
stimulates the money flow within the market network in the direction
opposite to the material flow, i.e. in the direction from the consumers to
the producers of raw materials. The conservation of money at the nodes
enables them to play a role of a certain aggregated information of the state
of the consumer medium as well as of the firm activity. The matter is that
for a certain collection of firms, e.g. firms, belonging to branch $i$, to
be able to supply firms of the lower rank linked directly with the given
firms of the network with the required input it is necessary and sufficient
that these firms possess the information characterizing the consumer state
in the region controlled by the given firms as a whole. Such information is
directly reflected in the price of the output.

A change in the consumer demand leads to variations of the material flow
within the market network. The latter, in turn, causes the firm's profits to
vary and, thus, the firms to increase or decrease their activities. In
particular, there are no barriers to the entry of new firms in respect to
the short-run profits being made in the given market. When the competition
is perfect this process will cause the average profit at each branch to be
maintained at zero value \cite{ec2,ec3}.

In the present Section within the framework of the market with a perfect
competition%
\index{perfect competition} we show that there may be certain hierarchical
structures%
\index{hierarchical structure} arising spontaneously which supply consumers%
\index{consumer} with goods ideally. This means that at the first
approximation the change in demand at one point of the consumer medium does
not cause variations in the goods flow at its other points although the
material flow varies across all branches belonging to all the hierarchy
levels.

Let us begin by setting up the model.

\subsection{Model}

\label{sa4.1}Let us consider an industry structure in the market involving
the consumer medium $\mathcal{M}$ and hierarchical network $\mathcal{N}$ of
the tree form supplying it with goods (Fig.~\ref{Fapp_2m}).

The material flow in the given network is determined by the collection $%
\{X_{i}\}$ of the total firm product $X_{i}$ at branch $i$ measured in mass
units. Because of the conservation of materials at each nodes, e.g. node $B$%
, we may write the expression 
\begin{equation}
(X_{i})_{in}=\sum_{j_{B}}(X_{j})_{%
\text{out}},  \label{1}
\end{equation}
where $(X_{i})_{in}$ and $(X_{j})_{\text{out}}$ are the total product at the
branches going in and out of the node $B$ and the sum runs over all the
branches leaving this node. In other words , the output $(X_{i})_{in}$ of
firm $i$ is equal to the sum of the inputs of the firms $(X_{j})_{\text{out}%
} $.

The branch $i$ is assumed to contain $n_{i}$ individual firms treated as
identical, at least, on the average. The total product $X_{i}$ at the branch 
$i$ is equal to the sum of the products $x_{i}$ produced by these $n_{i}$
firms, viz.:
\begin{equation}
X_{i}=n_{i}x_{i}.  \label{2}
\end{equation}
Variations in the number of firms $\{n_{i}\}$ is the market response to
change in the consumer demand%
\index{consumer demand}. The output collection $\{X_{i}\}$ of the firms
belonging to the last level of the network determines the corresponding set
of goods flows $\{X_{i}^{\ast }\}$ through the consumer medium: 
\begin{equation*}
X_{i}=X_{i}^{\ast }.
\end{equation*}

For the sake of simplicity we suppose that the $i$-th consumer is supplied
only by one of the firms at the last level. The latter can be justified, if
the consumers are substantially distinguished by there location in physical
or goods space.

The trade interaction at a node $B$ gives rise to a price $P_{B}$ for a mass
unit of the material exchanged in this interaction. As a result, the money
flow aggregated at branch $i$ due to trade interaction is $%
X_{i}(P_{i}^{(s)}-P_{i}^{(b)})$, where $P_{i}^{(s)}$, $P_{i}^{(b)}$
correspond to the nodes $B_{i}^{(s)}$, $B_{i}^{(b)}$ at which firms $i$ play
the role of a seller and buyer, respectively (Fig.~\ref{Fapp_2m}). The
individual profit of a firm belonging to branches $i$ is 
\begin{equation}
\pi _{i}=x_{i}(P_{i}^{\left( s\right) }-P_{i}^{(b)})-tc_{i}(x_{i}),
\label{3)}
\end{equation}
where the former term is its revenue and the latter one is its total cost%
\index{total cost} given by the expression \cite{ec2} 
\begin{equation}
tc_{i}(x_{i})=k_{i}+a_{i}x_{i}+b_{i}x_{i}^{2}.  \label{4}
\end{equation}
The set of parameters $\{k_{i},a_{i},b_{i}\}$ is considered to be constants
given beforehand. The total profit $\Pi _{i}$ at branch $i$ is 
\begin{equation*}
\Pi _{i}=X_{i}(P_{i}^{(s)}-P_{i}^{(b)})-[k_{i}n_{i}+a_{i}X_{i}+%
\frac{1}{n}_{i}b_{i}X_{i}^{2}].
\end{equation*}
The individual purpose of each firm is to maximize its profit with respect
to the product, which leads to the expression 
\begin{equation}
\frac{\partial \pi _{i}}{\partial x_{i}}=0,  \label{5}
\end{equation}
whence it follows that 
\begin{equation}
P_{i}^{(s)}-P_{i}^{(b)}=a_{i}+2b_{i}x_{i}.  \label{6}
\end{equation}

At the final stage firms sell their goods to ultimate consumers;
accordingly, the price $P_{i}^{(s)}$ is determined by the consumer market
demand. Assuming that at this final stage $P_{i}^{(s)}=d_{i}^{\ast
}-f_{i}^{\ast }X_{i}^{\ast }$, where $d_{i}^{\ast },f_{i}^{\ast }$ are
supposed to be predetermined constants and the firms transform each unit of
their input into one unit of the output after a processing cost of $%
c_{i}^{\ast }$ per unit is incurred \cite{ec2} we obtain. 
\begin{equation}
\Pi _{i}^{\ast }(X_{i}^{\ast })=X_{i}^{\ast }(d_{i}^{\ast }-f_{i}^{\ast
}X_{i}^{\ast })-P_{i}^{(b)}X_{i}^{\ast }-c_{i}^{\ast }X_{i}^{\ast }
\label{7}
\end{equation}

The condition of the profit $\Pi _{i}^{\ast }(X_{i}^{\ast })$ attaining
maximum with respect to $X_{i}^{\ast }$ leads to the expression 
\begin{equation}
P_{i}^{(b)}=d_{i}^{\ast }-c_{i}^{\ast }-2f_{i}^{\ast }X_{i}^{\ast }
\label{8}
\end{equation}
Perfect competition in the market maintains the profit $\pi _{i}(x_{i})$ at
the zero value \cite{ec3}, thus, for each firm of branch $i$ of the network
it can be written 
\begin{equation}
\left. \pi _{i}(x_{i}\left| P_{B}\right. )\right| =0  \label{9}
\end{equation}
for the $\{x_{i}\}$ and $\{P_{B}\}$ related to each other by expressions (%
\ref{6}), (\ref{8}).

Expression (\ref{6}), (\ref{8}) together with the conservation (\ref{1}) of
products at the nodes establishes such prices at the nodes that meet the
maximum profit condition for all the firms and satisfy the market
equilibrium.

Now let us analyse the characteristic properties of the stated model.

\FRAME{ftbpFU}{2.4595in}{2.8055in}{0pt}{\Qcb{Industry structure of the tree
form.}}{\Qlb{Fapp_2m}}{Fapp_2m}{\special{language "Scientific Word";type
"GRAPHIC";maintain-aspect-ratio TRUE;display "USEDEF";valid_file "F";width
2.4595in;height 2.8055in;depth 0pt;original-width 7.2964in;original-height
8.3333in;cropleft "0";croptop "1";cropright "1";cropbottom "0";filename
'Fapp_2m.gif';file-properties "XNPEU";}}

\subsection{Perfect self-regulation}

\label{sa4.2}

Substituting (\ref{6}) into (\ref{9}) we find that 
\begin{equation}
P_{i}^{(s)}-P_{i}^{(b)}=a_{i}+2\sqrt{k_{i}b_{i}}  \label{10}
\end{equation}
This result shows that the difference $P_{i}^{(s)}-P_{i}^{(b)}$ for each
branch $i$ is actually determined by the internal parameters of the
technology, production efficiency, and the market rate of the capital rather
then, by the consumer demand. The latter follows directly from a perfect
competition. This property enables us to find the price at a given node.
Setting for the sake of simplicity for firms belonging to the stem the buyer
price equal to zero (or ignoring it) we get for a node $B$ 
\begin{equation}
P_{B}=\sum_{i\in \mathcal{P}_{B}}\left( a_{i}+2\sqrt{k_{i}b_{i}}\right)
\label{11}
\end{equation}
where $\mathcal{P}_{B}$ is a path on the network connecting the given node $%
B $ with the stem.

Returning to the initial network shown in Fig.\ref{Fapp_1m}, we can
replicate the same speculations regarding with respect to the material flow
distribution $\{X_{i}\}$. In this way we will again obtain formula (\ref{11}%
) where, however, the path $\mathcal{P}_{B}$ is not unique. Therefore,
except for the degenerate cases, under a perfect competition the firms that
belong to the paths (fine lines on Fig.~\ref{Fapp_1m}) with larger values of
$P_{B}$ have to leave the market. The latter will convert the initial
production network of a complex geometry into a hierarchical network of the
tree form (solid line in Fig.~\ref{Fapp_1m}) which minimizes the price.

According to (\ref{8}) and (\ref{11}) the goods flow $X_{i}^{\ast }$
demanded by consumer $i$ can be given as 
\begin{equation}
X_{i}^{\ast }=\frac{1}{2f_{i}^{\ast }}[d_{i}^{\ast }-c_{i}^{\ast
}-\sum_{j\in \mathcal{P}_{i}}(a_{j}+2\sqrt{k_{j}b_{j}})]  \label{12}
\end{equation}
where $\mathcal{P}_{i}$ is the path leading from the stem to the given point 
$i$ of the consumer medium. The state of the consumer medium is entirely
specified by the set of parameters $\{d_{i}^{\ast },f_{i}^{\ast }\}$. Thus,
the change in the consumer demand reflects in time variations of the
parameters $\{d_{i}^{\ast },f_{i}^{\ast }\}$. Assuming the technological
parameters $\{a_{i},k_{i},b_{i},c_{i}^{\ast }\}$ of the production to be
constant it can be seen from (\ref{12}) that the goods flow $X_{i}^{\ast }$
through any given point $i$ of the medium $\mathcal{M}$ is controlled solely
by its own parameters $\{d_{i}^{\ast },f_{i}^{\ast }\}$. Time variations of
the parameters $\{d_{i}^{\ast },f_{i}^{\ast }\}$ at other points do not
affect the goods flow at the given point $i$. This property can be naturally
treated as perfect self-regulation.

Conservation of materials at the nodes (\ref{1}) allow us to find the
material flow $X_{i}$ going through a given branch $i$ as a function of the
goods flow through the consumer medium. Specifically, 
\begin{equation}
X_{i}=\sum_{j\in \mathcal{M}_{i}}X_{j}^{\ast }  \label{13}
\end{equation}
where $\mathcal{M}_{i}$ is the consumer medium region where the goods flow
as a whole is directly controlled by the given branch (Fig.~\ref{Fapp_2m}).
In particular, expressions (\ref{6}), (\ref{10}), (\ref{13}) yields the
number $n_{i}$ of independent firms participating in the production of the
output $X_{i}$:
\begin{equation}
n_{i}=\sqrt{\frac{b_{i}}{k_{i}}}\sum_{j\in \mathcal{M}_{i}}\frac{1}{%
2f_{j}^{\ast }}\left\{ d_{j}^{\ast }-c_{j}^{\ast }-\sum_{j^{^{\prime }}\in 
\mathcal{P}_{j}}(a_{j^{^{\prime }}}+2\sqrt{k_{j^{^{\prime }}}b_{j^{^{\prime
}}}})\right\}  \label{14}
\end{equation}
The last expression reflects the ability of the market as a whole to respond
to changes in the consumer demand by an appropriate change in the number of
firms $n_{i}$.

An opinion exists that the market economies are based solely on spontaneous
trade interactions between buyers and sellers in contrast to the
centrally-planned economies which are organized hierarchically. In the
present Section we tried to show that ``free markets'' might also contain
complex hierarchical structures, which, at first, arise spontaneously, and
then, minimize efforts to satisfy the consumer demands. Such hierarchical
systems are likely to provide the unique feasibility of self-processing of
information on which products to be supplied, how much each of them to be
produced, and in what ways to distribute them. Different aspects of this
problem has been considered also in works \cite{ga1,ga2,ga3,lu7}.

Real markets are not, naturally, perfect, firms are not identical and etc.
This leads to violation of the market ideal self-regulation. Nevertheless,
we think the presented model may be useful in analyzing real processes in
market if used as the first approximation since it enables one to take
directly into account possible complex interactions between producers.

\section{Self--regulation of trophic flows on hierarchical networks}

\label{sa5}

Let us consider some problems in interaction of different species forming a
certain closed ecological system%
\index{ecological system}. One of the classic approaches to the dynamics of
such systems is the Lotka--Volterra model%
\index{Lotka--Volterra model}. This model characterizes a given species $i$
by the number $n_{i}$ of its members measured in biomass%
\index{biomass} units and treated the species interaction in terms of
feeding relationships where one species plays the role of predators or preys
with respect to another. In other words, time variations in the values $%
\{n_{i}\}$ of the given ecosystem involving $N$ species%
\index{species} are governed by the collection of equations 
\begin{equation}
\frac{\partial n_{i}}{\partial t}=n_{i}\left( k_{i}(n_{i})+\sum_{j=1,j\neq
i}^{N}a_{ij}n_{j}\right)  \label{a4.1}
\end{equation}
where $k_{i}(n_{i})$ is the intrinsic growth parameter of the species $i$
depending generally on $n_{i}$, the quantity $a_{ij}$ is the constant of the
trophic interaction between the species $i,j$ which takes a positive or
negative value providing the species $i$ plays the predator or prey role in
the $ij$--interaction, and the matrix $\left( a_{ij}\right) $ is
antisymmetric that is $a_{ij}=-a_{ji}$ for $i\neq j$ and $a_{ii}=0$.

\FRAME{ftbpFU}{8.3164cm}{7.2247cm}{0pt}{\Qcb{Linear trophic chain (solid
lines) and an additional chain (a thin line) joined at the point B.}}{\Qlb{%
Fapp4_1}}{Fapp4_1}{\special{language "Scientific Word";type
"GRAPHIC";maintain-aspect-ratio TRUE;display "USEDEF";valid_file "F";width
8.3164cm;height 7.2247cm;depth 0pt;original-width 9.6297in;original-height
8.3333in;cropleft "0";croptop "1.0025";cropright "1.0001";cropbottom
"0";filename 'Fapp4_1.gif';file-properties "XNPEU";}}

The architectonics of the ecological system determines the matrix $\left(
a_{ij}\right) ,$ and a linear trophic chain (\ref{Fapp4_1}) of $2N$ elements
is one of the widely studied models for which the constants $\{a_{ij}\}$ are
different from zero for the nearest neighbors only and $a_{i(i-1)}>0$. Let
us consider the steady state solution of the system of equations (\ref{a4.1}%
) assuming for the sake of simplicity $k_{1}=\alpha $, $k_{2N}=-\beta $, and
$k_{i}=0$ (for $i=2,3,...,2N-1$), where $\alpha \,$ and $\beta $ are certain
positive constants. Omitting trivial mathematical manipulations we get for
the even terms $i=2,4,...,2N$%
\begin{equation}
n_{2}=\frac{\alpha }{a_{21}},\quad n_{i}=\frac{a_{(i-1)(i-2)}}{a_{i(i-1)}}%
n_{i-2}  \label{a4.2}
\end{equation}
and for odd ones $i=2N-1,2N-3,...,1$%
\begin{equation}
n_{2N-1}=\frac{\beta }{a_{2N(2N-1)}},\quad n_{i}=\frac{a_{(i+2)(i+1)}}{%
a_{(i+1)i}}n_{i+2}.  \label{a4.3}
\end{equation}
Whence, it follows that the population $n_{i}$ of the even species is
actually directly determined by the generation rate $\alpha $ of the initial
food source, whereas that of the odd species is controlled by the life time
of the last predators. The higher the generation rate, the greater the total
population of the ecosystem which reflects the fact of its adaptation to
variations in the environment. In particular, the population $n_{2N}$ of the
last species also increases as the generation rate $\alpha $ grows.

If there is another trophic chain%
\index{trophic chain} connected with the former one at the point \emph{B}
then, the two chains may be considered independently of each other, since,
at the point \emph{B} partial biomass flows $J$, $J^{\prime }$ are
controlled by the interaction of the last and the last but one elements
individually for each chain. So, in the general case such trophic chains
have to come into conflict with each other causing one of them to vanish
under stationary conditions. Indeed, otherwise according to (\ref{a4.2}) the
population $n_{2N}$ of the last species would be determined by the different
biomass flows $J$, $J^{\prime }$ at the same time. In other words, for
example, an increase in the generation rate $\alpha $ of the initial food
for the former chain will increase the population of the last level
predators, which in their turn will lead to local growth of the flow $%
J^{\prime }$ causing finally the latter chain to vanish as a whole. As a
rule, such a behavior of interacting trophic chains is also the case in the
dynamics when their populations vary in time (\ref{a4.1}).

However, the existence of a huge amount of species in nature and the growth
of their variety as one goes from higher to lower organisms show that there
must be a certain mechanism which, in principle, can stabilize ecosystems
and enable them to form hierarchically organized structures.

In the present Section we discuss a possible mechanism for trophic networks%
\index{trophic networks} of the tree form that prevents from preys being
killed by the predators whose population increases because of the growth of
a biomass flow in the neighboring branch of the trophic network.

\FRAME{ftbpFU}{1.7331in}{2.6377in}{0pt}{\Qcb{Hierarchically organized
irregular trophic network.}}{\Qlb{Fapp4_5}}{Fapp4_5}{\special{language
"Scientific Word";type "GRAPHIC";maintain-aspect-ratio TRUE;display
"USEDEF";valid_file "F";width 1.7331in;height 2.6377in;depth
0pt;original-width 5.4544in;original-height 8.3333in;cropleft "0";croptop
"1";cropright "1";cropbottom "0";filename 'Fapp4_5.gif';file-properties
"XNPEU";}}

First, we consider biomass flows on the trophic tree shown in Fig.~\ref
{Fapp4_5} whose distribution is controlled by the classic Lotka--Volterra
model with the coefficients $k_{i}=0$ for all the species except the species
of the last level feeding on the basic medium \QTR{cal}{M} of initial food
and the predators forming the tree stem. For the former and latter we set $%
k_{i}=\alpha _{i}$ and $k_{\text{\textrm{stem}}}=-\beta $, where $\{\alpha
_{i}\}$ and $\beta $ are certain positive constants. Going along the trophic
tree from the medium \QTR{cal}{M} to the stem we meet alternately species
whose population is actually directly determined by the state $\{\alpha
_{i}\}$ of the medium \QTR{cal}{M} and species whose population may be
treated as a free variable until we reach the stem or a branching point. In
Fig.~\ref{Fapp4_5} the former and the latter are labelled by the symbols ``%
\emph{m}'' and ``\emph{s}'', respectively. In mathematical terms the
movement from the species of the last level to the stem and in the opposite
direction is associated with solving the set of equations~(\ref{Fapp4_1}) by
the successive iteration. The main problem of this procedure occurs when we
meet a node where, as it has been shown for linear chains, the equations can
be incompatible. There are three types of nodes represented in Fig.~\ref
{Fapp4_5} by the branching points \emph{A}, \emph{B}, and \emph{C}. Let us
analyse their properties individually.

The simplest situation is realized at the point \emph{A}, where the branches
entering this point are actually independent of each other, the branch,
going out of the point \emph{A}, is of the ``\emph{s}'' type, and the
corresponding species have a net food source of the fixed rate of biomass
generation $\alpha _s=a_{s,m_1}n_{m_1}+a_{s,m_2}n_{m_2}$ formed by the
species $m_1$, $m_2$. Passing such a node we cannot meet any problem in
iterating equations~(\ref{a4.1}) until we reach another node.

At the point \emph{B} iterating equations~(\ref{a4.1}) is also associated
with no difficulty. The steady state conditions for the preys ``\emph{s}$%
_{1} $'' interacting with the predators at the point \emph{B} determine the
population of these predators ``\emph{m}$_{1}$'' which, therefore, must be
regarded as species of the ``\emph{m}'' type. The value $n_{m_{1}}$
determines the population $n_{s_{2}}$ of the species ``\emph{s}$_{2}$''. The
population $n_{s_{1}}$ of the species ``\emph{s}$_{1}$'' is a free variable
which is will be found by the following iterations.

Dealing with the point \emph{C} we meet the main problem. Going along
different branches ``\emph{s}$_{1}$'' and ``\emph{s}$_{2}$'' and entering
this node we have to ascribe two different values of the population $n_{m}$
to the same species ``\emph{m}'' what is impossible in the general case and
so, one of the branches ``\emph{s}$_{1}$'', ``\emph{s}$_{2}$'' will be wiped
out.

In order to avoid such a trouble we proposed the following self--regulation
model. Let us, first, single out one branching point on the trophic tree and
consider the interaction between its elements (\ref{Fapp4_2}). According to

\FRAME{ftbpFU}{3.3389cm}{3.4443cm}{0pt}{\Qcb{The node connecting one type of
predator with two types of prays.}}{\Qlb{Fapp4_2}}{Fapp4_2}{\special%
{language "Scientific Word";type "GRAPHIC";maintain-aspect-ratio
TRUE;display "USEDEF";valid_file "F";width 3.3389cm;height 3.4443cm;depth
0pt;original-width 4.4071in;original-height 4.5463in;cropleft "0";croptop
"1.0004";cropright "1.0002";cropbottom "0";filename
'Fapp4_2.gif';file-properties "XNPEU";}}

(\ref{a4.1}) the interaction between the predators $i$ and the preys $i_{1}$%
, $i_{2}$ gives rise to the partial biomass flows $J_{1}$, $J_{2}$ going
through the point \emph{C} determined by the expressions 
\begin{equation}
J_{1}=a_{1}nn_{1},\quad J_{2}=a_{2}nn_{2},  \label{a4.4}
\end{equation}
where $n$, $n_{1}$, $n_{2}$ are the populations of the predators and the
corresponding preys. The total biomass flow $J\,$ entering the species of
the predators during their chase is 
\begin{equation}
J=J_{1}+J_{2}=(a_{1}n_{1}+a_{2}n_{2})n.  \label{a4.5}
\end{equation}
In the classic Lotka--Volterra model the interaction coefficients $a_{1}$, $%
a_{2}$ are treated as given constants. In the model we consider the
predators to be ``active''; they can change the strategy of chase in order
to maximize the total biomass flow $J$. In mathematical terms this means
that the interaction coefficients $a_{1}(\phi )$, $a_{2}(\phi )$ are certain
functions of the chase strategy $\phi $ (treated as an additional free
variable belonging to the interval $[0,1]$), and that the predators
``choose'' the chase strategy in such a way that the function $%
J(n,n_{1},n_{2},\phi )$ attains its maximum for fixed values of $n$, $n_{1}$%
, $n_{2}$, that is 
\begin{equation}
J=n\max_{\phi \in \lbrack 0,1]}\left[ a_{1}(\phi )n_{1}+a_{2}(\phi )n_{2}%
\right] .  \label{a4.6}
\end{equation}
The maxima of the functions $a_{1}(\phi )$, $a_{2}(\phi )$ are attained at
different values $\phi _{1}=0$ and $\phi _{2}=1$ of the variable $\phi $
which means that the chase specialization with respect to one preys reduces
the chase efficiency with the other preys. The general form of the function $%
a_{j}(\phi ,\phi _{j})$ is demonstrated in Fig.~\ref{Fapp4_3}.

\FRAME{ftbpFU}{5.5091cm}{5.4015cm}{0pt}{\Qcb{Schematic view of functions $%
a_{1}(\protect\phi ,\protect\phi _{1}),a_{2}(\protect\phi ,\protect\phi
_{2}) $.}}{\Qlb{Fapp4_3}}{Fapp4_3}{\special{language "Scientific Word";type
"GRAPHIC";maintain-aspect-ratio TRUE;display "USEDEF";valid_file "F";width
5.5091cm;height 5.4015cm;depth 0pt;original-width 5.7778in;original-height
5.6662in;cropleft "0";croptop "1.0006";cropright "1.0013";cropbottom
"0";filename 'Fapp4_3.gif';file-properties "XNPEU";}}

Condition~(\ref{a4.6}) imposed on the chase strategy leads to the dependence
$\phi =\phi (n_{1},n_{2})$ which in turn gives rise to the dependence of the
interaction coefficients on the prey populations, $a_{1}=a_{1}(n_{1},n_{2})$%
, $a_{2}=a_{2}(n_{1},n_{2})$. Thus, a decrease in the population of one
preys or an increase in the others will reduce the chase efficiency with
respect to the former preys, preventing them from disappearing. This is
actually the essence of the given self--regulation model.

In order to illustrate this effect we consider in detail the case when the
functions $a_1(\phi ,\phi _1)$, $a_2(\phi ,\phi _2)$ are of the form 
\begin{equation}
a_1(\phi ,\phi _1)=a_1^0\left[ 1-\phi ^2\right] ,\quad a_2(\phi ,\phi
_2)=a_2^0\left[ 1-(1-\phi )^2\right] ,  \label{a4.7}
\end{equation}
where the parameters $a_1^0$, $a_2^0$ are constant. The asymptotic behavior
of the functions $a_1(\phi )\rightarrow 0$ as $\phi \rightarrow \phi _2=1$
and $a_2(\phi )\rightarrow 0$ as $\phi \rightarrow \phi _1=0$ allows us to
treat such a regulation as ideal because under these conditions none of the
species can be wiped out. Substituting (\ref{a4.7}) into (\ref{a4.6}) and
maximizing the result we find 
\begin{equation}
\phi (n_1,n_2)=\frac{a_2^0n_2}{a_1^0n_1+a_2^0n_2}  \label{a4.8}
\end{equation}
and 
\begin{equation}
a_1(n_1,n_2)=a_1^0\frac{a_1^0n_1(a_1^0n_1+2a_2^0n_2)}{(a_1^0n_1+a_2^0n_2)^2}%
,\quad a_2(n_1,n_2)=a_2^0\frac{a_2^0n_2(2a_1^0n_1+a_2^0n_2)}{%
(a_1^0n_1+a_2^0n_2)^2}.  \label{a4.9}
\end{equation}
As it must $a_1(n_1,n_2)\rightarrow 0$ as $n_1\rightarrow 0$ and $%
a_2(n_1,n_2)\rightarrow 0$ as $n_2\rightarrow 0$.

In order to demonstrate this model we will show for the point \emph{C} in
Fig.~\ref{Fapp4_5} that from the standpoint of species ``\emph{m}'' the
branches ``\emph{s}$_{1}$'' and ``\emph{s}$_{2}$'' may be aggregated into
one branch connected with an effective feeding medium. Let $n$, $n_{1}$, and 
$n_{2}$ be the populations of species ``\emph{m}'', ``\emph{s}$_{1}$'', and
``\emph{s}$_{2}$''. Then, the conservation of \ \ \ a biomass in species ``%
\emph{s}$_{1}$'', ``\emph{s} $_{2}$'' gives the equations 
\begin{equation}
a_{1}(n_{1},n_{2})n=\alpha _{1,\quad }a_{2}(n_{1},n_{2})n=\alpha _{2}.
\label{a4.10}
\end{equation}
For their compatibility it is required that 
\begin{equation}
\frac{a_{1}(n_{1},n_{2})}{a_{2}(n_{1},n_{2})}=\frac{\alpha _{1}}{\alpha _{2}}%
.  \label{a4.11}
\end{equation}
Substituting (\ref{a4.9}) into (\ref{a4.11}) and solving the obtained
equation we find that the total biomass flow through the species ``\emph{m}%
'' is equal to
\begin{equation}
J=a_{\mathrm{eff}}nC=\alpha _{\mathrm{eff}}C,  \label{a4.12}
\end{equation}
where $C=n_{1}+n_{2}$ is the total population of preys being in contact with
the predators ``\emph{m}'' and the effective coefficients
\begin{equation}
a_{\mathrm{eff}}=\frac{\rho ^{2}+\rho +1}{\rho +1}\frac{a_{1}^{0}a_{2}^{0}}{%
a_{1}^{0}+\rho a_{2}^{0}},  \label{a4.13}
\end{equation}
\begin{equation}
\alpha _{\mathrm{eff}}=\frac{\alpha _{2}a_{1}^{0}+\rho \alpha _{1}a_{1}^{0}}{%
a_{1}^{0}+\rho a_{2}^{0}}.  \label{a4.14}
\end{equation}
and the constant
\begin{equation*}
\rho =\left( \frac{\alpha _{1}a_{2}^{0}}{\alpha _{2}a_{1}^{0}}-1\right) +%
\sqrt{\left( \frac{\alpha _{1}a_{2}^{0}}{\alpha _{2}a_{1}^{0}}-1\right) ^{2}+%
\frac{\alpha _{1}a_{2}^{0}}{\alpha _{2}a_{1}^{0}}}.
\end{equation*}
Expression~(\ref{a4.12}) allows us to regard the given fragment of the
trophic tree as a linear chain.

Concluding the present Section we would like to note that such an ``active''
trophic network can adapt to variations in the environments without any
dramatic reorganization. The behavior of real ecological systems is certain
to be much more complicated, nevertheless, it is likely that active behavior
of species not, only predators but also preys lead to extension of a species
variety in nature (see also \cite{ga1,ga5}).

%\section[{BIBLIOGRAPHY}]{}


\addcontentsline{toc}{section}{BIBLIOGRAPHY}

\begin{thebibliography}{999}
\markboth
{\sc Bibliography}{\sc Bibliography}
\bibitem{parti} I.A.Lubashevsky, V.V.Gafiychuk Mathematical description
of the heat transfer in living tissue (Part I).
(http://xxx.lanl.gov/abs/adap-org/9911001)

\bibitem{1}  Alatortsev,V.L., Lubashevskii, I.A. Influence of anisotropy of
the diffusion coefficient on the asymptotic behavior of a spatial atom
distribution in structures with complex geometry. Sov. Phys. J. (USA). -
\textbf{31}, N5, 1988, pp.347-350.

\bibitem{3}  Andrushkiw, R.I., Gafiychuk,V.V., Lubashevskii, I.A. Two
boundary model for freezing processes in living tissue in: Proceedings of
the 14th IMACS World Congress , March 1994. Atlanta. pp.123-126.

\bibitem{2}  An introduction to the Practical Aspects of Clinical
Hyperthermia, Field S.B., Hand I.W (London, Taylor \& Francis, 1990).

\bibitem{4}  Antonets, V.A., Antonets, M.A., and Shereshevsky, I.A. In:
Fractals in the Fundamental and Applied Sciences, edited by H.O.Peitgen,
J.M.Hennigues, and L.F.Penedo. (Elsevier Science Publishers, Amsterdam,
1991) p.17.

\bibitem{5}  Antonets, V.A., Antonets, M.A., and Shereshevsky, I.A.
Stochastic Dynamics of pattern formation, Research Rep. in Phys. Nonlinear
waves 3., 1990, pp.307-315.

\bibitem{6}  Baish, I.W., Ayyaswamy, P.S., Foster, K.R. Heat transport
Mechanisms in Vascular Tissues: A Model Comparison, Trans. ASME J. Biom.
Eng., \textbf{108}, N11, 1986, pp.324--331.

\bibitem{7}  Baish, I.W., Ayyaswamy, P.S., Foster, K.R. Small - Scale
Temperature fluctuations in Perfused Tissue During local Hyperthermia,
Trans. ASME J. of Biom. Eng. \textbf{108}, 1986, pp.246--250.

\bibitem{BP84}  Bakhvalov, N.S. and Panasenko, G.P. Homogeniesation approach
to periodic media. Mathematical problems in the composite material
mechanics. (Nauka, Moscow, 1984) (in Russian).

\bibitem{8}  Berezovskii, A.A. Two - dimensional model of the cryosurgery of
living tissue, in: Mathematical modelling of physical process.- Kiev,
Institute of Mathematics Academy of Sciences of Ukraine 1989, pp.14--38 (in
Russian).

\bibitem{ec3}  Binger, B.R., Hoffman, E. Microeconomics with Calculus. -
(Harper Collins Publishers ,1988).

\bibitem{app.5}  Boulding, K.E. The unimportance of energy, in Energetics
and Systems\textit{.} W.J.Mitsch, R.K.Ragade, R.W.Bosserman, and J.A.Dillon
Eds., Ann Arbor Science, Ann Arbor, Michigan, 1982, pp.101--108.

\bibitem{app.1}  Busacker, R.G., and Saaty, T.L. Graphs and Networks. An
Introduction with Applications. - (McGraw Hill Book Company, New York, 1971).

\bibitem{9}  Charny, C.K., Levin, R.L. Bioheat transfer in a Branching
Countercurrent Network During Hyperthermia, Trans. ASME. J. Biom. Eng. 
\textbf{111}, N11, 1989, pp.263--270.

\bibitem{10}  Charny, C.K.,Weinbaum, S., Levin, R.L. An evolution of the
Weinbaum - Jiji Bioheat equation for Normal and Hyperthermia Conditions.
Trans. ASME J. Biom. Eng. \textbf{112}, N2, pp.81--87.

\bibitem{11}  Chato, I.C. Fundamentals of Bioheat Transfer, in: Thermal
Dosimetry and Treatment Planning ed. by M.Gautheries (Springer, Berlin,
1990), pp.1--56.

\bibitem{12}  Chato, I.C. Heat transfer to blood vessels. Trans. ASME J.
Biom. Eng. \textbf{102}, 1980, pp.110--118.

\bibitem{13}  Chen, M.M., and Holmes, K.R. Microvascular contributions in
tissue heat transfer. Ann. N.Y. Acad. Sci., \textbf{335}, 1980, pp.137--154.

\bibitem{14}  Chen, M.M. In Physical aspects of hyperthermia, edited by
G.Nussbaum (Amer. Inst. Phys., New York, AAPM Monograph \textbf{8}, 1983),
p.549; in Heat transfer in medicine and biology, Anal. and Appl., edited by
A.Shitzer and R.C.Eberhart (Plenum, New York, \textbf{1}, 1985), p.153.

\bibitem{15}  Comini, G., and Gindice, S.Del. Thermal aspects of
cryosurgery. J. Heat. Tran\-sfer. 1976, pp.534--549.

\bibitem{16}  Consensus Report of the Comac - BME workshop on Modelling and
treatment Planning in Loganissi. Greece (190. COMAC - BME Hyperthermia
Bulletin \textbf{4}, 1990, pp.26--49).

\bibitem{17}  Cooper, T.E., and .Trezek, G.J. Mathematical predictions of
cryogenic lesion, in: Cryogenics in Surgery, H. von Leden and W.G.Cohan
(eds.), (Medical Examination Publishing Co., New York, 1971), pp.128--149.

\bibitem{19}  Crezee, J. Experimental verification of thermal models. -(
Utrecht: University Utrecht, Faculty Geneeskunde. 1993).

\bibitem{18}  Crezee, J., and Lagendijk, J.I.W. Experimental verification of
bioheat transfer theories: measurement of temperature profiles around large
artificial vessels in perfused tissue. Phys. Med. Biol. \textbf{35}. N7,
1990, pp.905--923.

\bibitem{ec1}  Dolan, E.G., Lindsey, D.E. Microeconomics. - (Dryden Press,
1990).

\bibitem{21}  Essam, J.W. Percolation Theory. Rep. Progr. Phys. \textbf{43},
1980, pp.833--912.

\bibitem{22}  Feller, W. An Introduction to Probability Theory and its
Applications (Wiley, New York, 1971), \textbf{1,2}.

\bibitem{23}  Friedman, A. Variational principles and free boundary
problems.( John Wiley and Sons Inc., 1982).

\bibitem{24}  Gafiychuk, V.V., Lubashevskii, I.A., Osipov, V.V. Interface
dynamics of pattern formation in free boundary systems. Kiev. Naukova dumka,
212 p. (1990), (in Russian).

\bibitem{ga0}  Gafiychuk, V.V., Lubashevskii, I.A. Variational
representation of the projection dynamics and random motion of highly
dissipative systems. J.Math.Phys. \textbf{36}. N10, 1995,pp.5735-5752.

\bibitem{ga01}  Gafiychuk,\ V.V., Lubashevskii, I.A. Variational principles
of dissipative systems. J. of Sov. Math. \textbf{67}, No.2, 1993,p. 2943-2949

\bibitem{ga6}  Gafiychuk, V.V., Lubashevskii.I.A. Analysis of dissipative
structures based on Gauss variational principle. Ukr.mat.Zh. \textbf{44},
No.9, p.1085-1091, (1992). Plenum PC (Translated from ukrayins`kyy
matematychny zhyrnal.v.44. \#9. pp.1186-1192).

\bibitem{ga1}  Gafiychuk, V.V., Lubashevsky, I.A., Ulanowicz, R.E.
Distributed Self-regulation Induced by Negative Feedbacks in Ecological and
Economic Systems. e-print. http://xxx.lanl.gov/abs/adap-org/9811001 (1998).

\bibitem{ga2}  Gafiychuk, V.V., Lubashevsky, I.A., Klimontovich,\ Yu.L.
Mechanism of Self-regulation in a Simple Model of Hierarchically Organized
Market. e - print. http://xxx.lanl.gov/abs/adap-org/9903004 (1999).

\bibitem{ga3}  Gafiychuk, V.V., Lubashevskii, I.A. On hierarchical
Structures Arising Spontaneously in Markets with Perfect Competition.
J.Env.Systems. \textbf{25}, No.2. pp.159-166, (1996-1997).

\bibitem{ga4}  Gafiychuk, V., Lubashevskii, I., Andrushkiw, R. Two boundary
model for freezing front propagation in biological tissue. e-print .
http://xxx.lanl.gov/abs/patt-sol/9808009 (1998).

\bibitem{ga5}  Gafiychuk, V.V., Ulanowicz, R.E. Self-development and
distributed self-regulation in dissipative networks. Ref.No.CBL 96-010,
Cheasapeake Biological Laboratory, Solomons, Maryland, 1996.

\bibitem{25}  Gage, A.A., Guest, K., Montes, M., Caruana, J.A., Whalen, D.A.
Effect of varying freezing and thawing rates in experimental cryosurgery,
Cryobiology \textbf{22}, 1985, pp.175--182.

\bibitem{26}  Gardiner, C.W. Handbook of Stochastic Methods (Springer
Verlag, Berlin Heidelberg, 1983).

\bibitem{27}  de Gennes, P.-G. Scaling Concepts in Polymer Physics (Cornell
university press, Ithaca and London, 1979) .

\bibitem{app.2}  Gnedenko, B.V., Kovalenko, H.H. Introduction to the Theory
of Queueing Systems. - Nauka, Moscow, 1987 (in Russian).

\bibitem{28}  Hahn, G.M. Hyperthermia and Cancer (Plenum, New York, 1982).

\bibitem{29}  Harrison, A. Solid State Theory (McGraw - Hill Book Company,
N.Y., - London, Toronto, 1970).

\bibitem{30}  Heat transfer in medicine and biology. Analysis and
applications, edited by A.Shitzer and R.C.Eberhart (Plenum, New York, 1985), 
\textbf{1,2}.

\bibitem{MH92}  Hossain, M. Turbulent transport of a passive-scalar field by
using a renormalization-group method. Phys.Rev. A46. N 12, 1992,
pp.7608--7613.

\bibitem{31}  Jiji, L.M., Weinbaum, S., Lemons, D.E. Theory and experiment
for the Effect of Vascular Microstructure on Surface Tissue Heat Transfer -
Part 2. Model Formulation and Solution. Trans. ASME J. Biom. Eng. \textbf{106%
}, N11, 1984, pp.331--341.

\bibitem{app.7}  Jorgensen, S.E., Integration of Ecosystem Theories: A
Pattern. (Kluwer Academic Publishers, 1992).

\bibitem{G}  Kaur, I., and Gust, W. Fundamentals of Grain and Interphase
Boundary Diffusion..( Ziegler, Shtudgart, 1989).

\bibitem{32}  Kessel, R.C., and Kardon, R.H. Tissues and organs: a text
atlas of scanning electron microscopy (W:H.Freeman, San Francisco, 1979).

\bibitem{YKK93}  Kimura, Y., Kraichnan, R.H. Statistics of an advected
passive scalar. Phys. Fluids.A \textbf{5}, N 9, 1993, p.p. 2264--2277.

\bibitem{33}  Kittel, C. Quantum Theory of solids. (John Willy, N.Y. -
London, 1963).

\bibitem{34}  Klinger, H.G. The description of Heat Transfer in Biological
Tissue Annals of N.Y. Academy of Science \textbf{335} , 1980, pp.133--136.

\bibitem{35}  Lagendijk, J.I.W. A new theory to calculate temperature
distributions in tissues or why the ``bioheat transfer'' equation does work.
in Hyperthermia Oncology. Ed. by I.Overgrad (London Taylor \& Francis 1984)
pp.507--510.

\bibitem{36}  Lagendijk, J.I.W. The influence of blood flow in large vessels
on the temperature distribution hyperthermia. Phys. Med. Biol. \textbf{27},
1982, pp.17--23.

\bibitem{37}  Lagendijk, J.I.W. Thermal models: principles and
implementation in An Introduction to the Practical aspects of Clinical
Hyperthermia ed. by S.B.Field and I.W.Hand (London. Taylor \& Francis,
1990). pp.478--512.

\bibitem{38}  Landau, L.D., Lifshiz, E.M \ Hydrodynamic. Course on
Theoretical Physics \textbf{VI}. M: Nauka.- 1988 (in Russian).

\bibitem{40}  Lubashevskii, I.A., Alatortsev, V.L. Characteristics of
spatial distribution of diffusing atoms in regular polycrystalline samples.
Fiz. Met. Metalloved (USSR). - \textbf{65}, N5, 1988, pp.858-867 (in
Russian) (English transl. in: Phys. Met. Metall. (UK)).

\bibitem{41}  Lubashevskii, I.A., Alatortsev, V.L., and Keijan, A.G. Grain
boundary random walks and diffusion in polycrystals. Physica A \textbf{193},
1993,pp.259--303.

\bibitem{42}  Lubashevskii, I.A., Cadjan, A.G. Characteristics of
spatiotemporal fluctuation of temperature in living tissue. Phys.Rev.E 
\textbf{50}, 1994,pp. 2304-2307, .

\bibitem{ec4}  Lubashevskii, I.A., Gafiychuk, V.V. A simple model of
self-regulation in large natural hierarchical systems // J.Env.Syst. - 
\textbf{23(3)}, p.281-289 (1995).

\bibitem{lu0}  Lubashevsky, I.A., Gafiychuk, V.V. Mathematical model for a
perfect hierarchically organized system of life-support of distributed
living media, Proceedings of Russian Academy of Sciences, \textbf{351}, n.5,
pp. 611-613 (1996).

\bibitem{lu01}  Lubashevsky, I.A., Gafiychuk, V.V. Cooperative mechanism of
self-regulation in hierarchical living systems. e- print.
http://xxx.lanl.gov/abs/adap-org/9808003 (1998): to be published in SIAM
J.Appl.Math.

\bibitem{43}  Lubashevskii, I.A., Gafiychuk, V.V. Mathematical Modelling of
Heat and Mass Transfer in Living tissue. in Proceedings of the IMACS
Symposium on Mathematical Modeling, February \textbf{2--4}, 1994. Vienna,
pp.356--359.

\bibitem{39}  Lubashevskii, I.A. On the Theory of diffusion near single
dislocation. Chemical Physics (in Russian), \textbf{9}, N2, pp.272--286,
(1990).

\bibitem{lu1}  Lubashevsky,\ I.A., Gafiychuk, V.V., Priezzhev A.V. Effective
interface dynamics of heat diffusion limited thermal coagulation.
J.Biomed.Opt. v. 3, n. 1, pp. 102-111, (1998).

\bibitem{lu2}  Lubashevsky, I.A., Priezzhev, A.V., Gafiychuk, V.V., Cadjan,
M.G. Local thermal coagulation due to laser-tissue interaction as
irreversible phase transition. Journal of Biomed. Opt. \textbf{2}, \#1, pp.
95-105, (1997).

\bibitem{lu3}  Lubashevskii, I.A., Gafiychuk, V.V., Priezzhev, A.V.
Free-boundary model for local thermal coagulation. Growth of spherical and
cylindrical necrosis domains. Laser-Tissue Interaction VIII, S.L.Jacques,
Editor, Proc. SPIE, (1997).

\bibitem{lu4}  Lubashevsky, I.A., Gafiychuk,\ V.V. Bioheat transfer problem
and applications. In: Proceedings IMACS Symposium on Mathematical modeling,
Ed. I.Troch, F.Breitencker, February 5-7, Vienna, pp.949-955, (1997).

\bibitem{lu5}  Lubashevsky, I.A., Priezzhev, A.V., Gafiychuk,\ V.V., and
Cadjan, M.G. Dynamic free boundary model for laser thermal tissue
coagulation, In: Laser-Tissue Interaction and Tissue optics II, Proc. SPIE
v.2923, pp.48-57, (1996).

\bibitem{lu6}  Lubashevsky, I.A., Priezzhev, A.V., Gafiychuk, V.V. and
Cadjan, M.G. Free-boundary model for local thermal coagulation. Laser-Tissue
Interaction VII, S.L.Jacques, Editor, Proc. SPIE, v. 2681, p. 81-91, (1996).

\bibitem{lu7}  Lubashevsky, I.A., Gafiychuk, V.V., Klimantowich, Yu.L. Model
for a hierarchically organized market functioning perfectly. Mathematical
Modeling (in Russian, Moscow) \textbf{9}, \#5, pp.3-16 (1997).

\bibitem{BPC82}  Macroscopic properties of disorded media, edited by
R.Burridge, G.Papanicolaou and S.Childress. Lecture notes in physics. N 154
(Springer--Verlag, Berlin, 1982).

\bibitem{Pop2}  The Mathematics of Microcirculation Phenomena. J.F.Gross,
and A.S. Popel, eds. New York: Raven Press, (1980).

\bibitem{44}  Mchedlishvili, G.I. Microcirculation of Blood. General
Principles of Control and Disturbances. (Nauka Publishers, Leningrad, 1989)
(in Russian).

\bibitem{Pop1}  Microvascular Networks: Experimental and Theoretical
Studies. A.S.Popel, and P.C.Johnson, eds. Basel: S. Karger, (1986).

\bibitem{45}  Mooibroek, J., Lagendijk, J.J.W. A Fast Simple Algorithm for
the Calculation of Convective Heat Transfer by Large Vessels in There -
Dimensional in homogeneous Tissues. IEEE Transactions on Biomedical
Engineering. \textbf{38}. N5, pp.490--501, (1991).

\bibitem{46}  Natochin, Y.V. in: Human Physiology, edited by G.I.Kositsky
(Mir Publishers, Moscow, 1990) \textbf{2}, p.280.

\bibitem{app.9}  Nicolis, G., Prigogine, I. Self-organization in
nonequilibrium systems\textit{.(} Wiley, New York, 1977).

\bibitem{app.4}  Odum,H.T. Environment, Power and Society. (John Wiley and
Sons, New York, 1971).

\bibitem{OYS90}  Oleinik, O.A., Yosifian, G.A., and Shamaev, A.S.
Mathematical Problems in the Theory of Highly Heterogeneous Elastic Media
(Moscow State University Publishers, Moscow, 1990) (in Russian).

\bibitem{PO94}  Olla, P. Conteminent diffusion in a random array of fixed
parallel cylinders of high Reynolds numbers. Phys.Rev. E 50 , N 3, p.
2100--2108, (1994).

\bibitem{47}  Pennes, H.H. Analysis of tissue and arterial blood
temperatures in the resting human forearm. I. Appl. Phys., \textbf{1}, 93,
1948, pp.93--122.

\bibitem{48}  Physical Aspects of Hyperthermia, edited by G.H.Nussbaum,
(Amer. Inst. Phys., New York, AAPM Monograph \textbf{8}, 1982).

\bibitem{49}  Physics of Metals. I Electrons., edited by J.M.Ziman
(Cambridge, University press, 1987).

\bibitem{app.3}  Reading in Microeconomics. W.Breit, H.M.Hodrman,
E.Saueracker Eds., (Times Mirror / Mosby College Publishing, 1986).

\bibitem{50}  Roemer, R.B. Thermal Dosimetry, in Thermal Dosimetry and
Treatment Planning, ed. by M.Gautherie (Springer, Berlin, 1990), pp.119--214.

\bibitem{Rot}  Rothschild, M. Bionomics. The Instability of Capitalism.
(H.Holt and Company, Inc., 1990).

\bibitem{RHL77}  Rouche, N., Habets, P., and Laloy, M. Stability Theory by
Liapunov`s Direct Method. (Springer--Verlag, New York, 1977).

\bibitem{app.8}  Samuelson, P.A. AnalysFoundations of Economicis\textit{.(}
Cambridge: Harvard University Press, 1947).

\bibitem{S-P80}  Sanches--Palencia, E. Non--homogeneous Media and Vibration
Theory. Lecture notes in physics. N 127 (Springer--Verlag, Berlin, 1980).

\bibitem{ec2}  Scherer, F.M., Ross, D. Industrial Market Structure and
Economic Performance.( Houghton Mifflin Company. Boston , 1990).

\bibitem{51}  Shitzer, A., and Eberhart, R.C. in: Heat Transfer in Medicine
and Biology, Anal. and Appl., edited by A.Shitzer and R.C.Eberhart (Plenumm,
New York, 1985), \textbf{1}, p.137.

\bibitem{52}  Shoshenko, K.A., Golub, A.G., and Brod, V.I. Architectonics of
Circulation bed (Nauka Publishers, Novosibirsk, 1982) (in Russian).

\bibitem{53}  Solomon, A. On the use of mathematical models in cryosurgery,
Computer Science Div., Oak Ridge National Laboratory, TN, (1980).

\bibitem{55}  Song,W.J. et al. A combined Macro and Microvascular Model for
Whole Limb Heat Trans. J. Biom. Eng. \textbf{110}, N11, 1988, pp.259--268.

\bibitem{54}  Song, C.W., Lokshina, A., Rhee, I.G., Patten, M., and Levitt,
S.H. Implication of Blood Flow in Hyperthermia Treatment of Tumors. IEEE
Trans. Biom. Eng., \textbf{BME-31}, 1984, N1, pp.9--15 and the references in
it.

\bibitem{56}  Thermal Dosimetry and Treatment Planning, edited by
M.Gautherie (Springer, Berlin, 1990).

\bibitem{TL72}  Tennekes, H., Lumley, J.L. A first Course in Turbulence (MIT
Press,Cambridge. MA,1972).

\bibitem{LNP}  Trends and Applications of Pure Mathematics to Mechanics.
Lecture notes in physics. N 195 (Springer--Verlag, Berlin, 1984).

\bibitem{app.6}  Ulanowicz, R.E. Growth and development: ecosystems
phenomenology. (Springer - Verlag, New York, 1986).

\bibitem{VGC91}  Velasco, R.M., Garcia Colin,L.S. Kinetic approach to
generalized hydrodynamics. Phys.Rev. A. \textbf{44}, N 8.

\bibitem{57}  Weinbaum, S., and Jiji, L.M. A new simplified bioheat equation
for the effect of blood flow on local average tissue temperature. Trans.
ASME J. Biom. Eng. \textbf{107}, 1985, pp.131--139.

\bibitem{59}  Weinbaum, S., Jiji, L.M. Discussion of papers by Wissler and
Baish et al. concerning the Weinbaum - Jiji bioheat equation. Trans. ASME J.
Biom. Eng. \textbf{109}, 1987, pp.234--237.

\bibitem{58}  Weinbaum, S., Jiji, L.M., Lemons, D.E. Theory and Experiment
for the effect of Vascular microstructure on surface tissue heat transfer -
Part I: Anatomical Foundation and Model Conceptualization. Trans. ASME J.
Biom. Eng. \textbf{106}, 1984, pp.321--330.

\bibitem{60}  Weinbaum, S., Jiji, L.M. The matching of thermal fields
surrounding countercurrent microvessels and the closure approximation in the
Weinbaum - Jiji equation. Trans. ASME J. Biom. Eng. \textbf{111}, 1989,
pp.271--275.

\bibitem{61}  Whitmore, R.L. Rheology of Circulation.( Pergamon Press,
London, 1968).

\bibitem{62}  Wissler, E.H. An analytical Solution Countercurrent Heat
Transfer Between Parallel vessels with a linear Axial temperature gradient.
Trans. ASME J. Biom. Eng. \textbf{110}, N8, 1988, pp.254--256.

\bibitem{63}  Wissler, E.H. Comments on the new bioheat transfer equation
proposed by Weinbaum and Jiji. Trans.ASME J.Biom.Eng. \textbf{109}, 1987,
pp.226--233.

\bibitem{64}  Wulf, M. The energy conservation equation for living tissue.
IEEE Trans. Biom. Eng., \textbf{BME-21}, 1974, pp.494--495.

\bibitem{JR92}  Yuan, Jian-Yang, and Ronis, D. Theory of fully developed
hydrodynamic turbulent flow: Application of renormalization group methods.
Phys.Rev .A.\textbf{45}.N 8, 1992, pp.5578-5595.

\bibitem{65}  Ziman, J.M. Principles of the theory of Solids (University
press, Cambridge, 1972).
\end{thebibliography}
\end{document}